\documentstyle[aps,twocolumn,prl,epsf]{revtex}

\begin{document}

\newcommand{\be}{\begin{equation}}
\newcommand{\ee}{\end{equation}}
\newcommand{\bn}{\begin{eqnarray}}
\newcommand{\en}{\end{eqnarray}}

\draft

\twocolumn[\hsize\textwidth\columnwidth\hsize\csname @twocolumnfalse\endcsname

\title{\large \bf
Cluster coherent potential approximation for electronic structure 
of disordered alloys}

\author{M. S. Laad\cite{add}$^{1}$ and L. Craco$^{2}$}

\address{
$^{1}$Department of Physics, Loughborough University, LE11 3TU, UK \\
$^{2}$Institut f\"ur Theoretische Physik, Universit\"at zu K\"oln,
77 Z\"ulpicher Strasse, D-50937 K\"oln, Germany} 

\date\today

\maketitle

\widetext

\begin{abstract}
We extend the single-site coherent potential approximation (CPA) to include 
the effects of non-local disorder correlations (alloy short-range order) on
the electronic structure of random alloy systems.  This is achieved by mapping
the original Anderson disorder problem to that of a selfconsistently embedded
cluster.  This cluster problem is then solved using the equations of motion 
technique.  The CPA is recovered for cluster size $N_{c}=1$, and the disorder
averaged density-of-states (DOS) is always positive definite.  Various new
features, compared to those observed in CPA, and related to repeated 
scattering on pairs of sites, reflecting the effect of SRO are clearly 
visible in the DOS. It is explicitly shown that the cluster-CPA method 
always yields positive-definite DOS.  Anderson localization effects have 
been investigated within this approach. In general, we find that Anderson 
localization sets in before band splitting occurs, and that increasing 
partial order drives a continuous transition from an Anderson insulator 
to an incoherent metal.  
\end{abstract}


]

\narrowtext

\section{INTRODUCTION}

The problem of atomic short-range order (SRO) and its effect on the character
of electronic dynamics has been studied for many years now.  It is relevant to 
the detailed understanding of the conditions under which a transition from 
metallic to an Anderson localized (AL), disordered insulator occurs with 
increasing disorder strength~\cite{PWA}. In the $d=\infty$ limit, the 
well-known coherent potential approximation (CPA)~\cite{VKE} provides the 
{\it exact} solution of this Anderson disorder problem.  However, by 
construction, CPA cannot access specific effects (quantum interference from 
short-ranged scattering potentials) leading to AL behavior.  Such an endeavor 
demands explicit incorporation of the {\it dynamical} effect of SRO on carrier 
propagation in a disordered system, and thus, a cluster generalization of CPA.
Such attempts have indeed been carried out~\cite{GON,Mills}, but are extremely 
cumbersome numerically.  Additionally, they do not always guarantee the 
correct analyticity properties of the Green's functions: the few which do 
succeed in this respect, like the travelling cluster approximation 
(TCA)~\cite{Mills} and the dynamical cluster approximation 
(DCA)~\cite{Jarrell} are extremely tedious technically.  Further, the study 
of the effects of short-range order (SRO) on carrier dynamics has, to our best
knowledge, never been attempted using these approaches. It is obvious that 
attempting to tackle the harder problem of atomic (or doping induced) SRO 
along with arbitrarily strong local electronic correlations, not to mention 
important aspects like multi-orbital character of realistic transition 
metal-based oxide systems, with these approaches would be extremely 
numerically time-consuming.  Given this, it is imperative to develop 
semi-analytical routes as far as possible, resulting in much better (and 
easier) numerical tractability.

On the other hand, many important results have been gleaned from 
field-theoretic studies of the Anderson transition.  Perturbative 
renormalization group (RG)~\cite{Abrahams} approaches and extensions thereof, 
work in the weak disorder regime, and are a priori inapplicable in the
non-perturbative regime where the Anderson-Mott MIT would be expected to 
occur in the $3d$ case.  As a result, well-defined precursors of the MIT 
are observed even at very high temperatures, as experimentally 
demonstrated~\cite{PLee} in many systems in the form of scaling behavior 
of various quantites, breakdown of Matthiessen rule and the Mooij correlation.
Thus, while the perturbative $(d+\epsilon)$~\cite{Abrahams} approaches have 
indeed provided wealth of information, such approaches are insufficient at 
strong coupling, which is precisely the regime of interest for doped TMO 
systems, as well as systems like strongly interacting two-dimensional 
electron systems (in $Si$ MOSFETs~\cite{Mosfet}) which have been found to 
undergo insulator-metal transitions.  This is because one is always {\it 
effectively} in the strong disorder regime in strongly correlated electronic 
systems, where the renormalized one-elecron band width is very small (caused 
by Hubbard band-narrowing) in the correlated metal (or Mott insulating states,
where the band splitting {\it a la} Mott-Hubbard mechanism occurs).  
Consideration of such cases is out of bounds with perturbative approaches, 
and this requires development of genuinely non-perturbative approaches which 
should be capable of:

$(i)$ extending the CPA to access Anderson localization effects, and,

$(ii)$ having sufficient flexibility to incorporate effects of Mott-Hubbard 
physics via dynamical mean-field (DMF) or cluster-DMF approaches.
 
 In this paper, we devise a new {\it cluster}-CPA technique that satisfies 
the above requirements.  It is extremely simple to implement (requires only 
the solution of $N_{c}$ coupled, non-linear equations for the Green's function
for a cluster with $N_{c}$ sites), captures the {\it intracluster} 
correlations exactly, and is suited to further improvements (larger cluster 
size, incorporation of Mott-Hubbard physics).  As we will show, it also 
reproduces the exact CPA limit~\cite{VKE} for the single-site cluster 
($d=\infty$).

The CPA is tailor-made to describe electron dynamics in a system with purely 
diagonal disorder when correlations between spatially separated disorder 
scatterers can be ignored, a situation which is formally exact in $d=\infty$. 
Extensions to include off-diagonal (hopping) disorder within the CPA 
framework have been proposed~\cite{GON} by a variety of authors. Here, we will
propose a different (related to non-local CPA) Green's function technique.
Using the EOM technique, we explicitly include {\it dynamical} effects of 
arbitrarily strong scattering from short-range correlated disorder potentials.
In the process, we will make explicit contact with the problem of Anderson 
localization in disordered systems.

\section{MODEL AND SOLUTION}

The first step is the construction of a suitable cluster model 
Hamiltonian incorporating {\it diagonal} disorder.  Motivated by results of
Ref.~\cite{Mukul}, we generalize the Anderson disorder problem to finite 
dimensions by mapping the full Anderson disorder model,

\be
\label{eq1}
H=-t\sum_{<i,j>}(c_{i}^{\dag}c_{j}+h.c) + \sum_{i}v_{i}n_{i} \;,
\ee
to an effective model of a cluster of $N_{c}$ sites embedded in
an effective (dynamical) bath with a {\it complex} self-energy (matrix of 
size $N_{c}^{2}$).  Here, we assume a binary alloy distribution for disorder,
$P(v_{i})=(1-x)\delta(v_{i}-v_{A}) + x\delta(v_{i}-v_{B})$, and further, that
$<v_{i}v_{j}>-<v_{i}><v_{j}>=f_{ij} \equiv C$, a constant parameter. Strictly 
speaking, the SRO encoded in $f_{ij}$ is a function of $x$, temperature and 
other variables depending on the specific physical situation under 
consideration, and in real materials, this dependence should be explicitly 
taken into account.

In contrast to CPA, the method described below is tailor made to capture the 
dynamical effects of repeated scattering from a cluster of sites, which are 
correlated in a manner described by $f_{ij}$ over the cluster length scale.  
To proceed, we start with the embedded cluster Hamiltonian,

\bn
\nonumber
H &=& -t\sum_{\alpha}(c_{0}^{\dag}c_{\alpha}+h.c) +
v\sum_{\alpha}x_{\alpha}n_{\alpha} \\
&+& t\sum_{k,\alpha}e^{ik.R_{\alpha}}(c_{\alpha}^{\dag}c_{k}+h.c) \;,
\label{eq2}
\en
where $0, \alpha=1,..,z$ denote a central site $0$ coupled (via $t$) to $z$
nearest neighbors on a $d=z/2$ dimensional lattice, and the last term 
describes the hybridization of the boundary of the chosen cluster with an 
effective medium (conduction electron bath function) that has to be 
selfconsistently determined by a suitable imbedding procedure. We describe 
the details in course of the derivation below.

Defining the diagonal and off-diagonal propagators on the cluster as
$G_{00}(\omega)=<c_{0};c_{0}^{\dag}>, 
G_{\alpha 0}(\omega)=<c_{\alpha};c_{0}^{\dag}>, 
G_{\alpha\alpha}(\omega)=<c_{\alpha};c_{\alpha}^{\dag}>$, we start with
the equation of motion (EOM) for $G_{00}(\omega)$:

\be
\omega G_{00}(\omega)=1+t\sum_{\alpha} G_{\alpha 0}(\omega)+
v<x_{0}c_{0};c_{0}^{\dag}> \;.
\label{eq3}
\ee

Notice the appearance of a higher-order GF on the rhs of Eq.~(\ref{eq3}). It 
is naturally interpreted as the probability amplitude for having an electron
at a site $0$ with disorder potential $v$. Its EOM reads,

\be
(\omega-v)<x_{0}c_{0};c_{0}^{\dag}>=<x_{0}>+
t\sum_{\alpha}<x_{0}c_{\alpha};c_{0}^{\dag}> \;.
\label{eq4}
\ee

The EOM for $G_{\alpha 0} (\omega)$ on the rhs of Eq.~(\ref{eq3}) reads,

\be
\omega G_{\alpha 0}(\omega)=v<x_{\alpha}c_{\alpha};c_{0}^{\dag}> 
+ tG_{00}(\omega) + t\sum_{j\ne\alpha}G_{j0}(\omega) \;,         
\label{eq5}
\ee
and in a way similar to that leading to Eq.~(\ref{eq4}), we obtain,

\be
(\omega-v)<x_{\alpha}c_{\alpha};c_{0}^{\dag}>=t<x_{\alpha}c_{0};c_{0}^{\dag}> 
+ t\sum_{j\ne\alpha}<x_{\alpha}c_{j};c_{0}^{\dag}> \;.
\label{eq6}
\ee

Continuing along with identical lines for the various higer-order GFs 
generated in Eqs.~(\ref{eq3}-\ref{eq5}) give us

\bn
\nonumber
\omega<x_{0}c_{\alpha};c_{0}^{\dag}> &=& 
v<x_{0}x_{\alpha}c_{\alpha};c_{0}^{\dag}> +t<x_{0}c_{0};c_{0}^{\dag}> \\
&+& t\sum_{j\ne\alpha}<x_{0}c_{j};c_{0}^{\dag}>\;,
\label{eq7}
\en

\be
\omega<x_{\alpha}c_{0};c_{0}^{\dag}>=<x_{\alpha}> 
+ v<x_{\alpha}x_{0}c_{0};c_{0}^{\dag}> 
+ t<x_{\alpha}c_{\alpha};c_{0}^{\dag}> \;,
\label{eq8}
\ee

\bn
\nonumber
(\omega-v)<x_{0}x_{\alpha}c_{\alpha};c_{0}^{\dag}> &=& 
t<x_{0}x_{\alpha}c_{0};c_{0}^{\dag}> \\
&+& t\sum_{j\ne\alpha}<x_{0}x_{\alpha}c_{j};c_{0}^{\dag}>
\label{eq9}
\en

and,

\be
(\omega-v)<x_{\alpha}x_{0}c_{0};c_{0}^{\dag}> = <x_{0}x_{\alpha}> 
+ t<x_{0}x_{\alpha}c_{\alpha};c_{0}^{\dag}> \;.
\label{eq10}
\ee

Finally,
\be
(\omega-\epsilon_{k})<A_{0\alpha}c_{k};c_{0}^{\dag}> = 
t_{k}<A_{0\alpha}c_{\alpha};c_{0}^{\dag}>;
\label{eq11}
\ee
where $A_{0\alpha}=1,x_{0},x_{\alpha},x_{0}x_{\alpha}$ for the various types 
of Green functions which couple the bath back to the cluster (see above), and
$<x_{0\alpha}> \equiv <x_{0}x_{\alpha}>$ is the non-local correlation function 
of the disorder potential over the cluster length scale (scaling like $1/d$ in
$d$ dimensions). For a single site cluster, we recover the {\it exact} CPA 
result using the EOM for $G_{00},G_{k0},<x_{0}c_{0};c_{0}^{\dag}>$ and 
$<x_{0}c_{k};c_{0}^{\dag}>$ only.  Indeed, the local Green's function at the 
site $0$ is easily seen to be,

\be
G_{00}(\omega)=\frac{1-<x_{0}>}{\omega-\Delta(\omega)} + 
\frac{<x_{0}>}{\omega-v-\Delta(\omega)} \;,
\label{eq12}
\ee
with $\Delta(\omega)=\sum_{k}\frac{|t_{k}|^{2}}{\omega-\Sigma(\omega)
-\epsilon_{k}}$, which is exactly the CPA result.

In our cluster generalization, after a long and somewhat tedious algebra,
we finally find,

\be
G_{00}(\omega)=\frac{1+v <x_{0}c_{0};c_{0}^{\dag}> 
+ (v/t)F_{2}(\omega)<x_{\alpha}c_{\alpha};c_{0}^{\dag}>}
{\omega-zF_{2}(\omega)}
\label{eq13}
\ee
and,

\be
G_{\alpha 0}(\omega)=\frac{1}{\omega-\Delta(\omega)}
\frac{v\omega <x_{\alpha}c_{\alpha};c_{0}^{\dag}>
+t(1+v<x_{0}c_{0};c_{0}^{\dag}>)}{\omega-zF_{2}(\omega)}
\label{eq14}
\ee
where,

\be
v<x_{0}c_{0};c_{0}^{\dag}>=\frac{v<x_{0}>+(v/t)^{2}<x_{0}x_{\alpha}>
\frac{F_{1}(\omega)F_{2}(\omega)}
{\omega-v-F_{1}(\omega)}}{\omega-v-F_{2}(\omega)}
\label{eq15}
\ee
and,

\be
<x_{\alpha}c_{\alpha};c_{0}^{\dag}>=t^{-1}F_{1}(\omega)
<x_{\alpha}c_{0};c_{0}^{\dag}> \;,
\label{eq16}
\ee
with

\be
<x_{\alpha}c_{0};c_{0}^{\dag}>=\frac{<x_{\alpha}>
-<x_{0}x_{\alpha}>}{\omega-F_{1}(\omega)}
+\frac{<x_{0}x_{\alpha}>}{\omega-v-F_{1}(\omega)}\;.
\label{eq17}
\ee

Here, $F_{1}(\omega) \equiv \frac{t^{2}}{\omega-v-\Delta(\omega)}$ and 
$F_{2}(\omega)\equiv \frac{t^{2}}{\omega-\Delta(\omega)}$.  Finally,
the bath function, $\Delta(\omega)$ is computed from the 
equation~\cite{HIII,CG}, 

\be
G_{00}(\omega)=\int_{-W}^{+W}\frac{\rho_{0}(\epsilon) 
d\epsilon}{G_{00}^{-1}(\omega)+\Delta(\omega)-\epsilon} \;.
\label{eq18}
\ee
with $\Delta(\omega)=t^{2}G_{00}(\omega)$ for the Bethe lattice.
  At this point, one can show that the cluster-CPA technique developed above 
always yields positive-definite local DOS, defined by 
$\rho(\omega)=-Im G_{00}(\omega)/\pi$.  To show this explicitly, we observe 
that $G_{00}(\omega)$ can be brought to a convenient mathematical form 
by simple algebraic manipulations

\bn
\nonumber
G_{00}(\omega) &=& 
\frac{1-<x_0>}{\omega-F_{2}(\omega)}+\frac{<x_0>}{\omega-v-F_{2}(\omega)} 
\\ \nonumber
&-& \left[ \frac{<x_{\alpha}>-<x_{0}x_{\alpha}>}{\omega-F_{2}(\omega)}
+ \frac{<x_{0}x_{\alpha}>}{\omega-v-F_{2}(\omega)} \right]\\
&+& \frac{<x_{\alpha}>-<x_{0}x_{\alpha}>}{\omega-F_{1}(\omega)} 
+\frac{<x_{0}x_{\alpha}>}{\omega-v-F_{1}(\omega)}
\en

The first step in the derivation is to notice that each of the numerators 
is always positive definite by definition.  Clearly, to show that 
$\rho(\omega)$ is always positive definite, we have now only to show that 
$Im \Delta(\omega)\le 0$.  From the EOM technique used above, 
$\Delta(\omega)=\sum_{k}\frac{t_{k}^{2}}{\omega-\epsilon_{k}}$.  A 
straightforward calculation shows that $Im \Delta(\omega)\le 0$ for 
any choice of the unperturbed DOS, $\rho_{0}(\epsilon)\ge 0$.  Substitution 
in $G_{00}(\omega)$ above immediately shows that the disorder averaged DOS 
is always positive definite (clearly, self-consistency does not modify 
this conclusion).

Few clarifications concerning the physical meaning of the set of equations
is in order at this point.  First, we notice that the carrier dynamics is 
an explicit function of the higher-order (in $1/d$) SRO correlator, 
$<x_{0}x_{\alpha}>$.  It is also easy to check that the system of equations 
are exact both in the band and the atomic limit, and the CPA result is readily
recovered for the single-site cluster.  To interpret the meaning of the bath 
function, $\Delta(\omega)$ in our approach, we begin by observing that one 
can view {\it any} selfconsistent cluster approximation as being a valid 
description in a regime with short-ranged order on the cluster length scale 
(analogous to the single site approximation being formally exact at mean field 
level). An exact solution of the problem implies consideration of an infinite 
cluster, and, of course, is an insoluble problem.  Our choice for 
$\Delta(\omega)$ above is then linked to the mathematical consideration of 
short-ranged correlations over the cluster length scale only, or, equivalently,
to the consideration of dynamical effects of repeated scattering by a cluster 
consisting of a central site plus its $z$ nearest neighbors only.  The effects
of non-local SRO appears explicitly in the bath function $\Delta(\omega)$ (i.e,
in Eqn.~(\ref{eq18})) via $G_{00}(\omega)$ as defined in Eqn.~(\ref{eq13}) 
with its explicit dependence on $f_{ij}$.  It follows that the approach 
describes carrier dynamics in a situation where the carrier mean free path 
is of the order of the size of the chosen cluster ($l \simeq a$, the lattice 
constant for our cluster) in a fully selfconsistent way (see below), one step
beyond the CPA where $l=0$.

It is interesting to notice that $G_{00}(\omega), G_{\alpha 0}(\omega)$ can
be (formally) analytically expressed in terms of the corresponding diagonal
and off-diagonal cluster self-energies 
$\Sigma_{00}(\omega),\Sigma_{\alpha 0}(\omega)$ for
$d=1,2,...,\infty$, as well as on certain special lattices.  The above set of 
equations then constitute a closed set of simultaneous non-linear equations 
for the two self-energies, and are solved 
self-consistently to yield the renormalized (by disorder) DOS at the
central site, $\rho (\omega)=-\frac{1}{\pi} {\rm Im} G_{00}(\omega)$.

The alloy correlation function (describing SRO) is given by 
$<x_{0}x_{\alpha}>=<x_{0}><x_{\alpha}>+C_{0\alpha}$ in the general case, with 
$C_{0\alpha}$ encoding complete information about order-disorder instabilities
in the alloy. It is important to notice that the {\it dynamical} effect of 
strong scattering by these short-ranged correlations ($C_{0\alpha}$) on the 
electronic self-energy is explicitly included within our formulation above.  
In particular, the electron can undergo repeated scattering on the atomic 
sites within the chosen cluster, and, depending upon the degree and character 
of SRO (see below), can be localized due to interference effects coming from 
repeated scattering from spatially separated centers; i.e, via Anderson 
localization. To address the issue of Anderson localization in our NLCPA 
scheme, we follow Economou {\it et al.}~\cite{Economou} and use the 
localization function defined by:

\be
L(\omega)=Kt \mid G_{00}(\omega)
-\frac{G_{0\alpha}(\omega)G_{\alpha 0}(\omega)}{G_{00}(\omega)} \mid \;,
\label{eq22}
\ee
where electronic eigenstates with energy $\omega$ satisfying:

\begin{itemize}
\item $L(\omega)>1$ define {\it extended} states,

\item $L(\omega)<1$ define Anderson localized states, and,

\item $L(\omega)=1$ defines the mobility edge.
\end{itemize}
Here, $K$ is the connectivity of the lattice.  The formalism developed above 
thus allows a complete determination of the Anderson transition and its
dependence on lattice structure, type of SRO, and band-filling.

\section{NUMERICAL RESULTS AND DISCUSSION}

In this section, we describe the results obtained from the numerical solution
of the self-consistent set of coupled nonlinear equations derived in the 
previous section.  Since we are interested in generic effects of atomic SRO on
carrier dynamics, we choose the semicircular unperturbed DOS
$\rho_{0}(E)=\frac{2}{\pi W^{2}}\sqrt{W^{2}-E^{2}}$ as an approximation to 
the actual DOS for a three dimensional cubic lattice~\cite{HIII}.  This leads 
to a considerable simplification in the numerics without affecting the generic 
features qualitatively. We work with $W=1.424~eV$ and study the fully 
renormalized DOS, $\rho(\omega)$ and $L(\omega)$ as functions of the alloy 
composition $y=(<x_{0}>/(1-<x_{0}>))$, the atomic SRO parameter $C_{0\alpha}$ 
and the disorder strength $v$ for a half-filled band.

\begin{figure}[htb]
\epsfxsize=3.4in
\epsffile{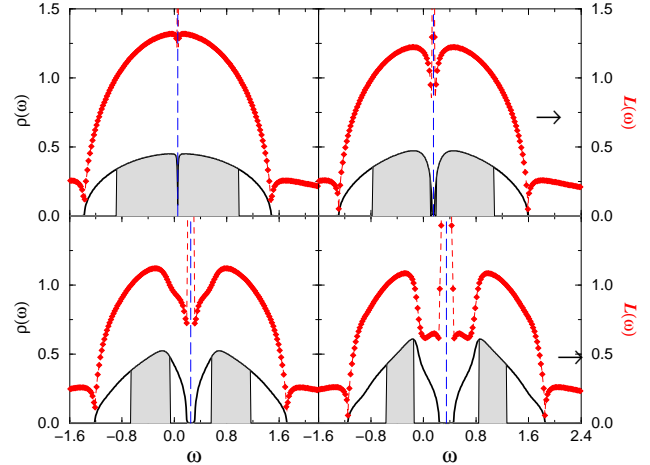}
\caption{Cluster-CPA DOS the binary-alloy distribution in 
the extreme SRO limit on a Bethe lattice for 
$<x_0>=0.5$, $<x_{0\alpha}>=0.0$ and various values of 
the local disorder potential, $v=0.1, 0.2, 0.3, 0.4$.  Shaded regions define
extended states, and unshaded regions define Anderson localized states.}
\label{fig1}
\end{figure}

We begin with the symmetric case with $y=1$, and extreme random disorder, 
i.e, $C_{0\alpha}=0$ or $<x_{0 \alpha}>=<x_{0}><x_{\alpha}>$. In $d=\infty$,
this corresponds to the CPA, with the metal-insulator transition occuring 
{\it continuously} at $v \ge W$.  Inclusion of SRO drastically changes the 
picture.  The M-I transition now occurs much earlier.  In fact, the band 
split regime occurs for $v/W \ge 1/4$.  However, states near and at the band 
center become Anderson localized {\it before} the band splits 
(Fig.~\ref{fig1}) and the metal (incoherent)- Anderson insulator transition 
is continuous. For $v < v_{c}$, the incoherent metal has a very 
similar character (breakdown of the quasiparticle) to that found in 
$d=\infty$ (CPA). It is also clear that the configuration averaged single 
particle DOS shows no anomalies across the Anderson localization transition, 
in agreement with well known~\cite{Dobr} arguments.   

\begin{figure}[htb]
\epsfxsize=3.5in
\epsffile{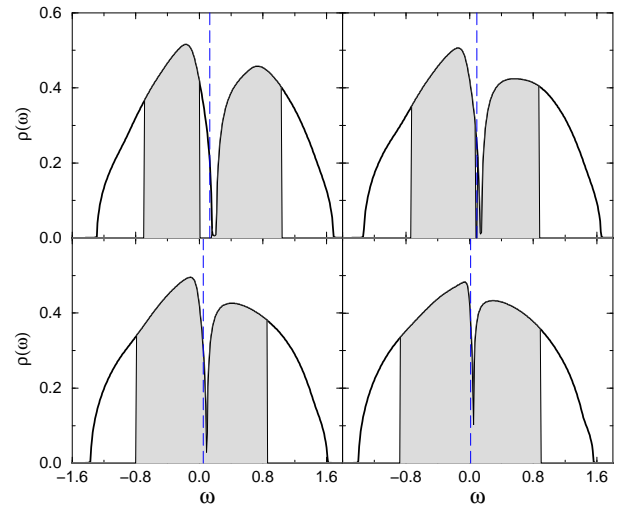}
\caption{(Cluster-CPA DOS on a Bethe lattice for 
$<x_{0\alpha}>=0.0$, $v=0.5$ and different alloy concentrations,
$<x_0>=0.4, 0.3, 0.2, 0.1$.}
\label{fig2}
\end{figure}

In Fig.~\ref{fig2}, we show the effect of changing the alloy composition on 
the critical value of $v$ needed to split the band.  For the case $y=2/3$ 
(i.e, $<x_{0}>=0.4$), a larger $v_{c}=0.5$ is required to localize sates 
at $E_{F}$, and it increases to $v_{c}=0.85$ for $y=3/7(x=0.3)$.  This 
Anderson insulating state (notice  that the Fermi level, denoted by the 
vertical lines in our plots, lies in the region of localized 
states) is explicitly related to our inclusion of the effect of carrier
scattering on short-ranged intersite atomic correlations $C_{0\alpha}=C$)
and is never observed in the CPA solution ($d=\infty$), which always predicts 
an incoherent metal for a (half-filled band) particle-hole asymmetric 
disorder distribution.  A continuous transition from an Anderson localized
insulator to an incoherent metal is clearly seen upon decreasing $y$ for 
a fixed disorder strength.  

\begin{figure}[htb]
\epsfxsize=3.5in
\epsffile{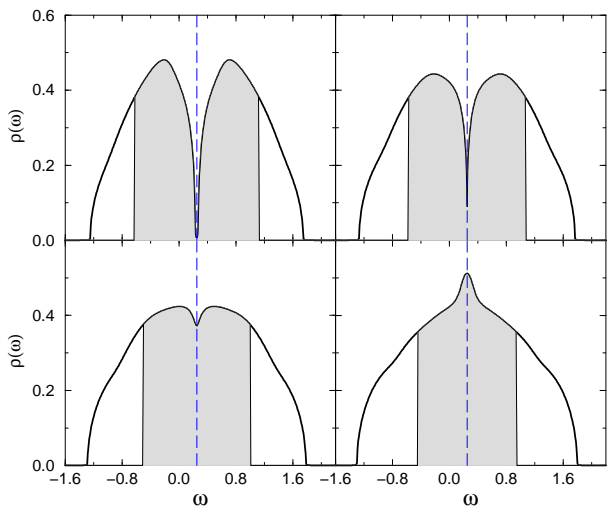}
\caption{Effect of partial SRO on the Cluster-CPA DOS for a Bethe lattice 
with $<x_{0}>=0.5$, $v=0.5$ and $<x_{0\alpha}>=0.1, 0.2, 0.3, 0.4$.}
\label{fig3}
\end{figure}

Next, we focus on the effect of varying the SRO parameter on the electronic 
structure.  In Fig.~\ref{fig3}, we show the DOS for $v=0.5$ for different 
values of $<x_{0 \alpha}>=0.1, 0.2, 0.3, 0.4$. 
Clearly, introducing partial order (actually, this corresponds to 
{\it increasing} $C_{0\alpha}$) results in increased tendency to itinerance, 
moving $E_{F}$ out of the region of localized states.  The AL insulator to
incoherent metal transition is clearly observed with increasing $C_{0\alpha}$,
and is a concrete illustration of an insulator-metal transition driven by the
degree of atomic SRO (partial order) in a system.  Clearly, increasing partial
order (notice that increasing $C_{0\alpha}$ corresponds to increasing the 
probability of having the {\it same} potential on the cluster sites) 
reduces the localizing effect of strong (repeated) intracluster disorder 
scattering, driving the I-M transition via increased itinerance. The 
situation is very analogous to the case where the pure Anderson disorder 
model is supplemented by additional short-range correlations in the 
hopping~\cite{Nature}, where increasing the off-diagonal randomness 
drives an insulator-metal transition for a fixed diagonal disorder strength.

\begin{figure}[htb]
\epsfxsize=3.5in
\epsffile{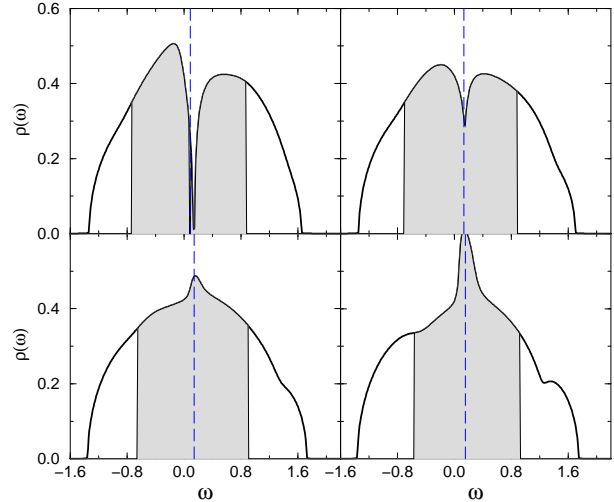}
\caption{Effect of partial SRO on the Cluster-CPA DOS for a Bethe lattice 
with $<x_{0}>=0.3$, $v=0.5$ and $<x_{0\alpha}>=0.0, 0.1, 0.2, 0.3$.}
\label{fig4}
\end{figure}

Qualitatively similar behavior is seen for an asymmetric alloy distribution.
In Fig.~\ref{fig4}, we show the DOS for $v=0.5$ and $y=3/7$ (i.e, 
$<x_{0}>=0.3$).  Interestingly, the spectrum shows additional features, but 
the AL insulator-metal transition with increasing $C_{0\alpha}$ follows the 
trend for the symmetric ($y=1$) case.

\begin{figure}[htb]
\epsfxsize=3.5in
\epsffile{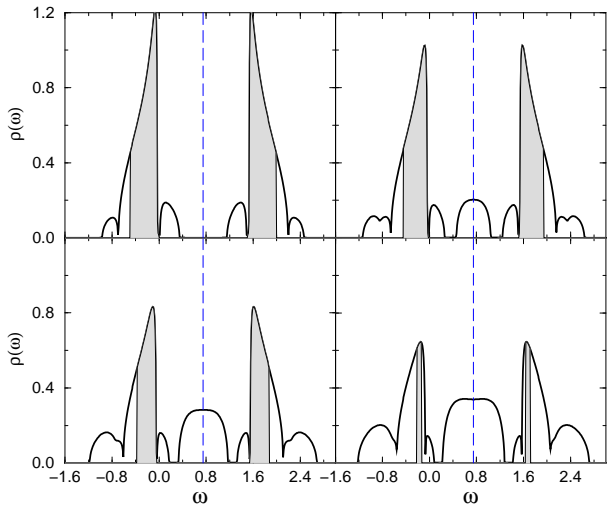}
\caption{Cluster-CPA DOS for $<x_{0}>=0.5$, $v=1.5$ and 
$<x_{0\alpha}>=0.0, 0.1, 0.2, 0.3$.}
\label{fig5}
\end{figure}

\begin{figure}[htb]
\epsfxsize=3.5in
\epsffile{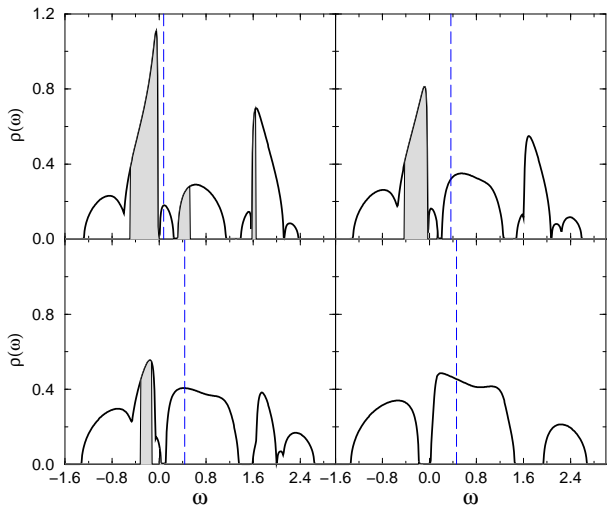}
\caption{Cluster-CPA DOS for $<x_{0}>=0.3$, $v=1.5$ and 
$<x_{0\alpha}>=0.0, 0.1, 0.2, 0.3$.}
\label{fig6}
\end{figure}

Increasing the ratio $v/W$ reveals rich structures in the 
DOS. In Fig.~\ref{fig5}, we show the one-electron DOS for $v=1.5$ with 
$<x_{0\alpha}>=0.0, 0.1, 0.2, 0.3$. Next, we turn our attention to 
Fig.~\ref{fig6}, which shows the evolution of the DOS for the asymmetric 
alloy distribution with $<x_{0}>=0.3$ for the same parameters. In this case, 
we are already in the split-band regime. Very rich structure is seen in the 
results. For comparison, we know that the corresponding DOS obtained for 
these cases within 
CPA ($d=\infty$, not shown) shows a split band structure with only upper- 
and lower ``Hubbard'' bands. Obviously, CPA is incapable of resolving the 
fine structure in the DOS originating from repeated scattering between 
spatially separated scattering centers.  The rich structures seen in the 
cluster generalization correspond partially to these effects, and can be 
traced back to the spectrum of eigenstates of the isolated cluster. In fact, 
the multiple sub-bands can easily be shown to be centered around 
eigen-energies of the isolated cluster for the case of the symmetric alloy 
distribution with $f_{0\alpha}=0$. However (Figs.~\ref{fig5}-\ref{fig6}), 
in the general case with $f_{0\alpha}\ne 0$,
one sees eight or nine distinct subband structures.  We interpret the 
additional structures as arising from atomic SRO (non-zero $f_{0\alpha}$);
in particular from ``shake-up'' effects originating from strong resonant 
scattering of carriers (from the atomic SRO) from cluster sites.  For 
comparison, we remark that coupling the two-site cluster to the ``bath'' 
(rest of the lattice) via second order processes in the hopping 
(corresponding to the ``Hubbard I'' approximation for the cluster) is 
incapable of accessing SRO effects in a consistent way.  In particular, in 
addition to violating the Hubbard sum rules~\cite{HI}, it cannot 
yield ``shake-up'' features in the DOS, always yielding {\it only} six 
bands centered around the eigenvalues of the 2-site cluster, and broadened 
by an amount $O(t)$. This discussion shows the importance of treating the 
effects of {\it both} itinerance (via $\Delta(\omega)$) and the (incoherent) 
resonant scattering on the same footing, and reveals the weaknesses inherent 
in uncontrolled approximations. 

Additional interesting features observed from the calculations deserve 
comments.  We clearly observe that the localization function $L(\omega)$
shows non-analytic behavior near each subband edge, but no 
non-analyticities across the mobility edge.  More detailed characterization 
of the Anderson insulator-metal transition requires a detailed study of the 
two-particle responses~\cite{DW}(density correlations, optical 
conductivity) and is left for future work.

Our results show some resemblance to those obtained by Rowlands 
{\it et al.}~\cite{Gyorffy} for the same model (in $1d$) using the 
Korringa-Kohn-Rostoker Non-Local Coherent Potential Approximation (NLCPA) 
employing the dynamical cluster approximation (DCA).  However, no attempt 
has been made there to study Anderson localization.  Further, our treatment 
of partial SRO is very different from theirs.  It is worth pointing out that 
our results, along with the NLCPA ones, are quite different from those 
obtained by Jarrell {\it et al.}~\cite{Jarrell}.  However, we are presently 
unable to quantify the reasons behind these differences.

We emphasize that the approach developed here has a wide applicability to 
various problems where the effect of atomic (chemical), magnetic, Jahn-Teller, 
etc., SRO on the character of carrier dynamics is an important issue.  In 
particular, it should be applicable to the problem of electronic structure 
of disordered TM alloys~\cite{TM}, and to more recent cases such as 
hole-doped manganites~\cite{EDagotto}, where a plethora of experimental work 
clearly demonstrates the importance of such effects in a correlated 
environment.  Applications to such systems is in progress and will be 
reported elsewhere. 
  
\acknowledgments
Work carried out (LC) under the auspices of the Sonderforschungsbereich 
608 of the Deutsche Forschungsgemeinschaft. MSL acknowledges the financial 
support from the MPIPKS, Dresden.

\end{document}